\documentclass[prd,superscriptaddress,amsfonts,amssymb,amsmath,showpacs,twocolumn]{revtex4-2}
\usepackage{bm}
\usepackage{amsfonts}
\usepackage{mathtools}

\usepackage{latexsym}
\usepackage[utf8]{inputenc}
\usepackage{graphicx}
\usepackage{amsmath}
\usepackage{palatino}
\usepackage{mathpazo}
\usepackage{textcomp}
\usepackage{mathrsfs}

\usepackage{float}
\usepackage{booktabs}
\usepackage{dcolumn}
\usepackage{multirow}
\usepackage{hyperref}
\hypersetup{colorlinks,citecolor=blue}
\usepackage{amsmath}
\usepackage{xcolor}
\usepackage{orcidlink}
\usepackage[caption=false]{subfig}
\usepackage{commath}
\def\jnl@style{\it}
\def\aaref@jnl#1{{\jnl@style#1}}

\def\aaref@jnl#1{{\jnl@style#1}}

\def\aj{\aaref@jnl{AJ}}                   % Astronomical Journal
\def\apj{\aaref@jnl{ApJ}}                 % Astrophysical Journal
\def\apjl{\aaref@jnl{ApJ}}                % Astrophysical Journal, Letters
\def\apjs{\aaref@jnl{ApJS}}               % Astrophysical Journal, Supplement
\def\apss{\aaref@jnl{Ap\&SS}}             % Astrophysics and Space Science
\def\aap{\aaref@jnl{A\&A}}                % Astronomy and Astrophysics
\def\aapr{\aaref@jnl{A\&A~Rev.}}          % Astronomy and Astrophysics Reviews
\def\aaps{\aaref@jnl{A\&AS}}              % Astronomy and Astrophysics, Supplement
\def\mnras{\aaref@jnl{Mon.~Not.~Roy.~Astron.~Soc.}}             % Monthly Notices of the RAS
\def\prd{\aaref@jnl{Phys.~Rev.~D}}        % Physical Review D
\def\prc{\aaref@jnl{Phys.~Rev.~C}}  % Physical Review C
\def\prl{\aaref@jnl{Phys.~Rev.~Lett.}}    % Physical Review Letters
\def\qjras{\aaref@jnl{QJRAS}}             % Quarterly Journal of the RAS
\def\skytel{\aaref@jnl{S\&T}}             % Sky and Telescope
\def\ssr{\aaref@jnl{Space~Sci.~Rev.}}     % Space Science Reviews
\def\zap{\aaref@jnl{ZAp}}                 % Zeitschrift fuer Astrophysik
\def\nat{\aaref@jnl{Nature}}              % Nature
\def\aplett{\aaref@jnl{Astrophys.~Lett.}} % Astrophysics Letters
\def\apspr{\aaref@jnl{Astrophys.~Space~Phys.~Res.}} % Astrophysics Space Physics Research
\def\physrep{\aaref@jnl{Phys.~Rep.}}      % Physics Reports
\def\physscr{\aaref@jnl{Phys.~Scr}}       % Physica Scripta
\def\commat{\aaref@jnl{Comm.~Math.~Phys.}}              % Communications in Mathematical Physics
\def\science{\aaref@jnl{Science}}               % Science
\def\cqg{\aaref@jnl{Classical Quant.~Grav.}}            % Classical and Quantum Gravity
\def\jpcs{\aaref@jnl{JPCS}}                                     % Journal of Physics Conference Series
\def\ijmpd{\aaref@jnl{Int.~J.~Mod.~Phys.~D}}                    % International Journal of Modern Physics D
\def\grg{\aaref@jnl{Gen.~Relat.~Gravit.}}               % General Relativity and Gravitation
\def\rpp{\aaref@jnl{Rep.~Prog.~Phys.}}          % Reports on Progress in Physics
\def\npa{\aaref@jnl{Nucl.~Phys.~A}}        % Nuclear Physics A
\def\lrr{\aaref@jnl{Living Rev.~Rel.}}                   % Living reviews in relativity
\def\jcap{\aaref@jnl{J.~Cosmology Astropart.~Phys.}}    % Journal of cosmology and astroparticle physics
\def\rmp{\aaref@jnl{Rev.~Mod.~Phys.}}   %Reviews of modern physics
\def\epjc{\aaref@jnl{Eur.~Phys.~J.~C}}

\allowdisplaybreaks[1]
\begin{document}

\color{black}       %% For one column

\title{Kuchowicz gravastars in the braneworld formalism}

\author{Oleksii Sokoliuk\orcidlink{0000-0003-4503-7272}}
\email{oleksii.sokoliuk@mao.kiev.ua}
\affiliation{Main Astronomical Observatory of the NAS of Ukraine (MAO NASU),\\
Kyiv, 03143, Ukraine}
\affiliation{Astronomical Observatory, Taras Shevchenko National University of Kyiv, \\
3 Observatorna
St., 04053 Kyiv, Ukraine}

\author{Alexander Baransky\orcidlink{0000-0002-9808-1990}}
\email{abransky@ukr.net}
\affiliation{Astronomical Observatory, Taras Shevchenko National University of Kyiv, \\
3 Observatorna
St., 04053 Kyiv, Ukraine}
\author{P.K. Sahoo\orcidlink{0000-0003-2130-8832}}
\email{pksahoo@hyderabad.bits-pilani.ac.in}
\affiliation{Department of Mathematics, Birla Institute of Technology and
Science-Pilani,\\ Hyderabad Campus, Hyderabad-500078, India.}

%%%%%%%%%%%%%%%%%%%%%%%%%%%%%%%%%%%%  DATE  %%%%%%%%%%%%%%%%%%%%%%%%%%%%%%%%%%%%
\date{\today}
\begin{abstract}

In the current letter, we study isotropic static spherically symmetric gravastars without charge under the framework of braneworld gravity (dimensionally reduced RS-II braneworld with positive brane tension) using the metric potential of Kuchowicz type (which is physically acceptable, non-singular and stable) in the Mazur-Motolla conjuncture. We derived the Kuchowicz free parameters from the junction conditions assuming Schwarzschild vacuum without cosmological constant. As well, we assume that the interior is filled with Dark Energy (DE-like) fluid, shell with ultrarelativistic stiff fluid. Adopting this conditions, we probed several physical aspects of the gravastars, such as proper length, shell energy and entropy, surface redshift and adiabatic index, interior region mass.

\end{abstract}

\maketitle
\section{Introduction} \label{sec:1}

Over the past decade, interest in gravastars has increased greatly. There was written a large number of papers, that investigate the research of charged and non-charged gravastars (for example, see \cite{Pani2010,Chan2010,Kubo2016,Banerjee2020,Ghosh2020,Shamir2020,Abbas2020,Kuhfittig2020}). Gravastars (gravitational condensate stars) was first ever proposed in works of Mazur and Mottola \cite{Mazur2001,Mazur2004}. This stars are probable alternative to the black holes. Gravastars are usually have the following structure:
\begin{itemize}
    \item Interior region $\mathcal{D}_1$ (from $r=0$ to $R_1$): de Sitter fluid with Equation of State (further - EoS) $p=-\rho$.
    \item Intermediate (shell) region $\mathcal{D}_2$ (from $R_1$ to $R_2$): stiff Zeldovich fluid with EoS $p=\rho$. 
    \item Exterior region $\mathcal{D}_3$ (from $R_2$ to $r=\infty$): empty spacetime with Schwarzschild, Schwarzschild-de Sitter or Reissner-Nordstr\"om geometry and EoS $p=\rho=0$. 
\end{itemize}
Exterior spacetime could be considered as the Schwarzschild Black Hole (BH) metric.

\begin{figure*}[!htbp]
    \centering
\begin{tikzpicture}[thick,scale=1.0,,
    mycirc/.style={circle,fill=red!70, minimum size=0.01cm}]

\draw (0,0) -- (1.5,-1.5);
\draw (3,0) -- (1.5,-1.5);

\draw[] (0,0) -- (0,-6);

\draw (1.5,-4.5) -- (0,-6);
\draw  (1.5,-4.5) -- (3,-6);

\draw (0,-3) -- (1.5,-1.5);

\draw (0,-3) -- (1.5,-4.5);
\draw (1.5,-4.5) -- (3,-3);

\draw (3,-3) -- (1.5,-1.5);

\node[rotate=90] at (-0.2,-3) 
    {\large $r=0$};
\node[rotate=45] at (0.7,-2) 
    {\large $r=R$};
\node[rotate=135] at (0.7,-4) 
    {\large $r=R$};

 \draw (1.5,-1.5) node[] (C) {};  
        \draw (1.5,-4.5) node[] (D) {};  
            \draw[dashed] (C) to [bend left=-35] (D);
 \draw (1.5,-1.5) node[] (C) {};  
        \draw (1.5,-4.5) node[] (D) {};  
            \draw[dashed] (C) to [bend left=35] (D);
 \draw (0,-3) node[] (C) {};  
        \draw (3,-3) node[] (D) {};  
            \draw[dashed] (C) to [bend left=35] (D);
 \draw (0,-3) node[] (C) {};  
        \draw (3,-3) node[] (D) {};  
            \draw[dashed] (C) to [bend left=-35] (D);
 \draw (0,-3) node[] (C) {};  
        \draw (0,-6) node[] (D) {};  
            \draw[dashed] (C) to [bend left=25] (D);
 \draw (0,0) node[] (C) {};  
        \draw (0,-3) node[] (D) {};  
            \draw[dashed] (C) to [bend left=25] (D);
\node[] at (3.2,0) 
    {\Large $i_0$};
    \node[] at (3.2,-3) 
    {\Large $i_0$};
        \node[] at (3.2,-6) 
    {\Large $i_0$};
        \node[] at (2.5,-1) 
    {\Large $\mathscr{I}_-$};
            \node[] at (2.5,-2) 
    {\Large $\mathscr{I}_+$};
                \node[] at (2.5,-4) 
    {\Large $\mathscr{I}_-$};
                    \node[] at (2.5,-5) 
    {\Large $\mathscr{I}_+$};
\end{tikzpicture}   
2\caption{Conformal diagram of static gravastar}
    \label{fig:1}
\end{figure*}
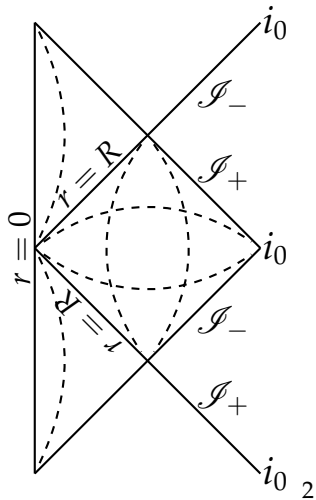

On the Figure (\ref{fig:1}) we illustrated the geometry of the gravastar spacetime on the conformal diagram. We expanded static gravastar spacetime as the Reissner-Nordstrom spacetime. Metric potentials of the gravastar interior spacetime are non-singular, and therefore, at $r=0$ we have space without any singularities. On the conformal diagram, triangles mean de Sitter (further - dS) spacetime (interior gravastar region with EoS parameter $\omega=-1$). As well, $i_0$ stands for the infinetly distant spacelike point and $\mathscr{I}_{+/-}$ stands for the null-like hypersurface.

Mainly, there was the research in gravastars field done only in the Einstein General Theory of Relativity. But, despite the fact that Einstein's relativism still describes the universe quite well, we cannot quantize relativistic systems, and recent cosmological observations and theoretical works require modifying the classical Einstein-Hilbert action. There was many attempts done to properly modify GR gravity, and one of the most viable theories of modified gravity is $f(\mathcal{R})$ theory, that replaces Ricci scalar in classical EH action by arbitrary function of Ricci scalar. This theory was originally proposed in \cite{Buchdahl1970}. One of the interesting features of this kind of modified gravity is that this MOG could describe cosmological inflation \cite{Brooker2016,Huang2013,Starobinsky1980} as well as the late time acceleration, solve the dark energy problem \cite{Capozziello2011,Nojiri2017}. There was done some research on the various topics in other kinds of MOG's. For example, Das et al \cite{Das2017} have derived exact solutions of gravastars in the $f(\mathcal{R},\mathcal{T})$ gravity. In this model, they defined pressure as the negative energy density, shell region was filled with the ultrarelativistic fluid and exterior region was assumed to be the vacuum in the non-rotating Schwarzschild-de Sitter spacetime (with the present $\Lambda$ term). Gravastar solutions in this gravity were non-singular and exact. In the $f(\mathcal{G},\mathcal{T})$ gravity, gravastar model was first ever constructed by Shamir et al. \cite{Shamir2020}. In turn, there was also probed the electromagnetic nature of the gravastars by \cite{Debnath2019} in $f(\mathcal{T})$ gravity with: $\mathcal{T}=0$ (traceless) or $f_{\mathcal{T}\mathcal{T}} = 0$. It was shown that with $\mathcal{T}=0$ there is no physically acceptable solutions, but with  $f_{\mathcal{T}\mathcal{T}} = 0$ authors constructed non-singular and exact solutions for three gravastar regions. 

As we know, gravity is much weaker than other three natural forces - strong and weak nuclear force and electromagnetic force. In the particle physics, this problem is known as the hierarchy problem. In attempt to solve this problem, Randall and Sundrum proposed RS-I model \cite{Randall1999,PhysRevLett.83.4690}. This model includes two $(3+1)$ dimensional branes with the positive and negative tension in the $5D$ bulk (usually Anti-de Sitter). In the RS braneworlds, only gravity could freely propagate through bulk, and other forces located on the branes. We will investigate the second model, namely RS-II, in which the second brane with negative tension is sent to the infinity, and thus we have only one positive tension brane. In this model at the lower energies, we could also recover Newtonian gravity. During the last few decades, many papers were devoted to the investigation of braneworlds, and some of them were particularly aimed on the gravastars (for example see \cite{Banerjee:2015ipa,Sengupta2020,Arbanil:2019xfi} and references therein. In our paper we will investigate the non-charged Kuchowicz gravastars in the framework of braneworld gravity, which is exactly the dimensionally reduced RS-II model with five dimensional bulk.

We have found some analytical and numerical solutions for different regions of gravastars structure. Also, we investigated the physical aspects of the gravastars, such as proper length, energy, entropy and interior region mass. 

Our letter is organised as follows: in the Section (\ref{sec:1}) we provide the brief introduction into the topic of charged and non-charged gravastars, modified theories of gravity. In the Section (\ref{sec:22}), we describe the formalism of braneworld gravity, derive energy density and pressure for the spherically symmetric interior spacetime from modified Einstein Field Equations. In the Section (\ref{sec:3}), we provide effective Equation of State for different gravastar regions and describe Kuchowicz-like metric potential that we will use across our paper. In the Section (\ref{sec:4}) we probe the physical aspects of the gravastars in the framework of modified gravity. Finally, we summed up everything in the last Section (\ref{sec:6}).

\section{Branewolrd formalism} \label{sec:22}

In braneworld theory of gravity the Einstein-Hilbert (EH) action integral is modified as follows \cite{Maartens:2003tw}:
\begin{equation}
\begin{gathered}
  S_\mathrm{BWG} =
  \frac{1}{2\kappa^2_{4+d}} \int d^4x d^dy\sqrt{-^{(4+d)}g}
  \left[^{(4+d)}R- 2\Lambda_{4+d}\right]\\
  + \frac{1}{2\kappa^2_{4}}\int d^4x\sqrt{-g}(-\sigma+\mathcal{L}_\mathrm{M})
 \end{gathered}
 \end{equation}
where $\mathcal{L}_\textrm{M}$ is the Lagrangian of the matter fields. Then, by varying the EH action we could obtain (modified) EFE \cite{Sotiriou2010}:
\begin{widetext}
\begin{equation}
\begin{gathered}
    {}^{(4+d)\!}G_{AB} \equiv  \;{}^{(4+d)\!}R_{AB}-{1\over2}
  \;{}^{(4+d)\!} R \;{}^{(4+d)\!}g_{AB} = -\Lambda_{4+d}
  \;{}^{(4+d)\!}g_{AB}+ \kappa_{4+d}^2 \;{}^{(4+d)\!}T_{AB}
\end{gathered}
\label{eq:13}
\end{equation}
\end{widetext}
where $4+d$ dimensional energy-momentum tensor is related to the $4$-dimensional on brane one by Dirac delta function:
\begin{equation}
    ^{(4+d)}T_{AB}=-\Lambda_{4+d} \;^{4+d}g+(-\sigma g_{\mu\nu}+T_{\mu\nu})\delta(y-y_0)
\end{equation}
In the equation above $y_0$ means the location of the brane in the additional fifth bulk coordinate $y$. As well, aforementioned EFE's for 5D Randall-Sundrum II braneworld configuration could be rewritten in the more simplified form \cite{Maartens:2003tw}
\begin{widetext}
\begin{eqnarray}
  {G}_{\mu\nu}&=&-{1\over2}{\Lambda}_5 g_{\mu\nu}+{2\over3}
  \kappa_5^2 \left[{}^{(5)}T_{AB}g_\mu{}^A g_\nu{}^B +
  \left( {}^{(5)}T_{AB}n^An^B-{1\over 4} \;{}^{(5)}T \right)
  g_{\mu\nu} \right]
  \nonumber \\
  && + K K_{\mu\nu}-K_\mu{}^\alpha K_{\alpha\nu} +
  {1\over2}\left[K^{\alpha\beta}K_{\alpha\beta}-K^2 \right]g_{\mu\nu} -
  {\cal E}_{\mu\nu},
  \label{ein}
\end{eqnarray}%
\end{widetext}
where four dimensional terms above could be defined through the five dimensional ones:
\begin{equation}
  {\cal E}_{\mu\nu} =
  {}^{(5)\!}C_{ACBD} \, n^Cn^D g_\mu{}^A g_\nu{}^B,
\end{equation}
\begin{equation}
    \Lambda=\frac{1}{2}(\Lambda_5+\kappa^2_4\sigma)\xRightarrow{\Lambda=0}\Lambda_5=-\kappa^2_4\sigma
\end{equation}
\begin{equation}
    \kappa_4^2=\frac{1}{6}\lambda\kappa_5^4
\end{equation}
Finally, we as well could derive curvature from the Israel–Darmois junction conditions:
\begin{equation}
    K_{\mu\nu}=-\frac{1}{2}\kappa_5^2\bigg[T_{\mu\nu}+\frac{1}{3}(\sigma-T)g_{\mu\nu}\bigg]
\end{equation}
After the use of some tedious algebra and Israel's junction conditions, we could finally come up with the most simplified form of the on brane field equations \cite{PhysRevD.62.024012}:
\begin{equation}
\begin{gathered}
    G_{\mu\nu} = T_{\mu\nu}+\frac{6}{\sigma}S_{\mu\nu}+E_{\mu\nu}
    \end{gathered}
    \end{equation}
where the expressions for unknown tensors are
\begin{equation} \begin{gathered}
    S_{\mu\nu} = \frac{TT_{\mu\nu}}{12}-\frac{T_{\mu\alpha}T^{\alpha}_{\nu}}{4}+\frac{g_{\mu\nu}}{24}(3T_{\alpha\beta}T^{\alpha\beta}-T^2)
    \end{gathered}
 \end{equation}
 \begin{equation}
\begin{gathered}
    E_{\mu\nu} = -\frac{6}{\sigma}\bigg[Uu_{\mu}u_\nu+P\chi_\nu \chi_\nu + h_{\mu\nu}\bigg(\frac{U-P}{3}\bigg)\bigg]
    \end{gathered}
    \end{equation}
In the equations above, $\sigma$ is the $(3+1)$
dimensional brane tension, $G_{\mu\nu}$ is the Einstein tensor,
$T_{\mu\nu}$ is the brane stress-energy tensor ($T=g^{\mu\nu}T_{\mu\nu}$
is it's trace), $U$ and $P$ are bulk energy density and isotropic
pressure, finally $u_\mu$ is the four-velocity and
$\chi_\nu = 1/\sqrt{g_{rr}}\delta^\nu_r$ is radial spacelike unitary
vector, $h_{\mu\nu}=g_{\mu\nu}+u_{\mu}u_\nu$. For simplicity, we will
use bulk Equation of State (EoS) $P=\omega U$ with $U=A\rho+B$ where
$\rho$ is the brane energy density. We will study the $(3+1)$ dimensional gravastar geometry with the following interior spherically symmetric spacetime (metric signature is $(-,+,+,+)$):
\begin{equation}
    ds^2 = -e^{\nu(r)}dt^2 + e^{\lambda(r)}dr^2 + r^2d\theta^2+r^2\sin^2\theta d\phi^2
    \label{eq:2.3}
\end{equation}
Using the line element above and EFE's from Equation (\ref{eq:13}), we could derive energy density and isotropic pressure \cite{Sengupta2020}:
\begin{equation}
e^{-\lambda}\left(\frac{\lambda'}{r}-\frac{1}{r^2}\right)+\frac{1}{r^2}
 =\left[\rho(r)  \left( 1+\frac {\rho(r) }{2 \sigma} \right) +{\frac {6 U}{\sigma}}\right],\label{eq6}
 \end{equation}
 \begin{widetext}
 \begin{equation}
e^{-\lambda}\left(\frac{\nu'}{r}+\frac{1}{r^2}\right) -\frac{1}{r^2},
 =\left[p \left( r \right) +{\frac {\rho \left( r \right)  \left( p \left( r \right) +\frac{\rho(r)}{2} \right)}
{\sigma}}+{\frac {2U}{\sigma}}+{\frac {4 P}{\sigma}}\right],\label{eq7}
\end{equation}
\begin{equation}
e^{-\lambda}\left[\frac{\nu''}{2}-\frac{\lambda' \nu'}{4}+\frac{\nu'^2}{4}+\frac{\nu'-\lambda'}{2r}\right] = \Bigg[p(r)+{\frac{\rho(r) \left(p(r)+\frac{\rho(r)}{2}\right)}{\sigma}}+{\frac{2U}{\sigma}}-{\frac {{2 P}}{\sigma}}\Bigg].\label{eq8}
\end{equation}
\end{widetext}
where we have used stress-energy tensor of form \cite{Shamir2020}:
\begin{equation}
    T_{\mu\nu}=(\rho+p_t)u_\mu u_\nu -p_tg_{\mu\nu}+(p_r-p_t)\chi_\mu \chi_\nu
\end{equation}
Here, we define $p_r$ and $p_t$ as radial and tangential pressures respectively, $u_\mu$ is timelike four-velocity and $
\chi_\mu$ is radial four-vector. We consider isotropic case for simplicity and therefore $p_r=p_t=p$. It is also necessary to define 'effective' energy density and isotropic pressure \cite{Arbanil:2019xfi}
\begin{equation}
    \rho^{\mathrm{eff}}=\rho\bigg(1+\frac{\rho}{2\sigma}\bigg)+\frac{6U}{\sigma}
\end{equation}
\begin{equation}
    p^{\mathrm{eff}}=p+\frac{1}{2\sigma}\bigg[\rho(\rho+2p)+4U\bigg]
\end{equation}
Finally, in braneoworld gravity for a given line element, energy conservation 
(well known Tolman-Oppenheimer-Volkov) equation reads \cite{Oppenheimer1939,Poncede1993,Rahaman2014,Tolman1939}:
\begin{equation}
    \frac{dp}{dr}+\frac{\nu'(r)}{2}(\rho+p)+ F_{\mathrm{ex}} = 0
    \label{eq:7}
\end{equation}
where $F_{\mathrm{ex}}$ is the external force, that is present because of stress-energy tensor non-continuity in braneworld MOG. 
To properly analyze such compact astrophysical object as gravastar, one could assume the physically viable metric potential, namely Kuchowicz-like metric potential of form \cite{Kuchowicz1968}:
\begin{equation}
    e^{\nu(r)} = e^{Cr^2+2\ln D} 
\end{equation}
where $C$ and $D$ are arbitrary constants. Interior spacetime with the given forms of metric potentials is often called as Kuchowicz spacetime.

\subsection{Junction conditions}

Gravastars need to satisfy continuity equations below (equations obtained at the surface of gravastar of radius $R$, therefore $r=R$) \cite{BHAR2021100879}:
\begin{equation}
    \mathrm{Continuity\;of\;}g_{tt}:\quad 1-\frac{2M}{R}=e^{CR^2}D^2
\end{equation}
\begin{equation}
    \mathrm{Continuity\;of\;}\frac{\partial g_{tt}}{\partial r}:\quad \frac{2M}{R^2}=2CRe^{CR^2}D^2
\end{equation}
So the solutions of equations above for Kuchowicz constants are
\begin{equation}
    C=-\frac{M}{R^2 (2 M-R)}
\end{equation}
\begin{equation}
    D=\frac{\sqrt{M} e^{-\frac{C R^2}{2}}}{\sqrt{C} R^{3/2}}
\end{equation}
Therefore, we could proceed to the gravastars in the next section.

\section{Gravastars in the braneworld gravity} \label{sec:3}
\subsection{Interior region}

Spacetime of gravastars is separated on three different regions. First region - interior region.  Fluid in this region has the following effective Equation of State (further - EoS) \cite{Mazur2001,Mazur2004}:
\begin{equation}
    p = -\rho
\end{equation}
Also, for the interior region one equation is true \cite{Abbas2020,Sharif2020}:
\begin{equation}
    p = -\rho = - \rho_c
\end{equation}
where $\rho_c$ is constant energy density.
Then, by adopting the effective EoS $p=-\rho_c$, we could derive metric tensor component (metric potential). 
For five dimensional RS-II braneworld, gravastar with K metric potential has the energy density of form
\begin{equation}
e^{-\lambda}\bigg(\frac{\lambda^{\prime}}{r}-\frac{1}{r^2}\bigg)+\frac{1}{r^2}=
\bigg[\rho_c\bigg(1+\frac{\rho_c}{2\sigma}\bigg)+\frac{6}{\sigma}(A\rho_c+B)\bigg]
\end{equation}
\begin{widetext}
\begin{equation}
e^{-\lambda}\bigg(\frac{1}{r^2}+\frac{\nu^{\prime}}{r}\bigg)-\frac{1}{r^2}=
\bigg[-\rho_c\bigg(1+\frac{\rho_c}{2\sigma}\bigg)+\frac{2}{\sigma}(A\rho_c+B)+\frac{4\omega}{\sigma}(A\rho_c+B)\bigg]
\end{equation}
\end{widetext}
Therefore, one left metric potential has the following solution:
\begin{equation}
  \begin{gathered}
      e^{-\lambda}=-\frac{6 A \rho_c r^3+6 B r^3+\rho_c r^3
   (\rho_c+\sigma )-3 r \sigma +3 c_1}{3 r \sigma }
  \end{gathered}
\end{equation}
To obtain regular solution at the origin we impose $c_1=0$ so that
\begin{equation}
    e^{-\lambda}=1-\frac{r^2 (\rho_c(6 A+\rho_c+\sigma )+6
   B)}{3 \sigma }
\end{equation}
Now we could also derive the brane tension from continuity condition for $g_{rr}$:
\begin{equation}
    \mathrm{Continuity\;of\;}g_{rr}:\quad \bigg(\frac{1-2M}{R}\bigg)^{-1}=e^{\lambda(R)}
\end{equation}
Using equation above, brane tension is
\begin{equation}
    \sigma =\frac{R^3 (\rho_c (12 A+\rho_c)+12 B)}{12 M-2 \rho_c R^3}
\end{equation}
For the above expression, brane tension is positive (as expected), if $A>0$ for relatively small constant density $\rho_c\ll1$ and if $A<0$ for bigger values of $\rho_c$.
Using that assumption for brane tension we finally obtain simplified form of metric potential
\begin{equation}
    e^{-\lambda}=1-\frac{2 M r^2}{R^3}
\end{equation}
Also, we could see that both energy density and isotropic pressure do not suffer from the central singularity, which is common for gravastars models with non-singular metric potentials.
Consequently, while we already defined stress-energy tensor components, we finally could also calculate the total interior region mass from the formula \cite{Ghosh2021}:
\begin{equation}
    \mathcal{M} = \int^{R_1=R}_0 4\pi r^2 \rho dr=\frac{4\pi R^3}{3}\rho_c
    \label{eq:18}
\end{equation}
As we see, interior region mass is invariant under the change of gravitation framework, since energy density is constant. But we also could define the active gravitational mass of interior region in terms of effective energy density as follows:
\begin{equation}
    \widetilde{M}= \int^{R_1=R}_0 4\pi r^2 \rho^{\mathrm{eff}} dr=\frac{4}{3} \pi  R^3 \bigg(\frac{6 A \rho_c}{\sigma }+\rho_c \bigg(\frac{\rho_c}{2 \sigma }+1\bigg)\bigg)
\end{equation}

We plot the gravitational mass for both regular and effective energy densities on the Figure (\ref{fig:1}). As we see, quantity of the active gravitational mass grows exponentially as we get nearer to the envelope (shell), which is expected (for example, the same results were obtained for gravastars, admitting conformal motion in $f(R,T^2)$ gravity \cite{SHARIF2021}).
\begin{figure}[!htbp]
    \centering
    \includegraphics[width=0.7\columnwidth]{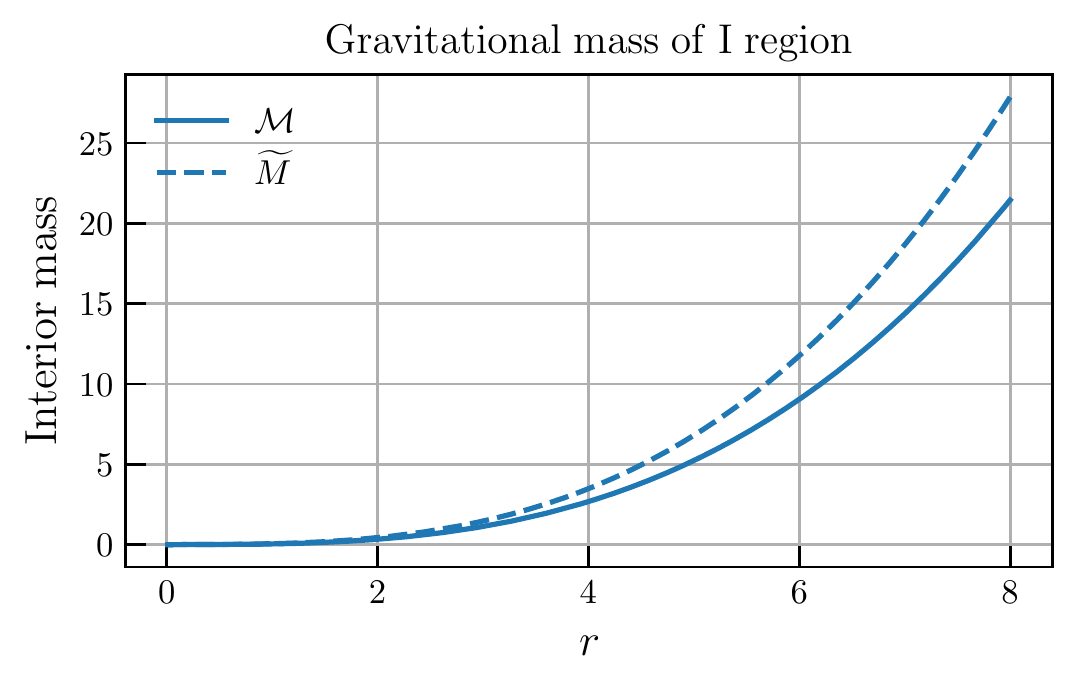}
    \caption{Active gravitational mass of the first (interior) region for PSR J1416-2230 compact star with mass $M=1.97M_\odot$ and radius $R=9.69$km \cite{Demorest2010ATN}. Since constant density in the interior region is considered to be relatively small ($\rho_c=0.01$), we assume that $A=1$, and therefore $\sigma=20.0762$.}
    \label{fig:12}
\end{figure}

\subsection{Intermediate region: shell}

The shell of the gravastar is usually very thin, but finite. It separate interior and exterior regions of gravastar and contain all of the collapsing star mass. We will assume, that the matter in the shell obey EoS equation $\rho=p$ (with $\omega=1$). Also, from the thin-shell approximation, $0<e^{-\lambda(r)}<1$ \cite{Abbas2020}.Thus, with the given EoS, we could say the fluid in the shell is stiff fluid (found by \cite{Zeldovich1972}). For stiff fluid-like equation of state general form of braneworld EFE's are rewritten as follows \cite{Sengupta2020}
\begin{equation} \label{eq16}
\frac{e^{-\lambda}\lambda^{\prime}}{r}+\frac{1}{r^2}=\bigg[\rho\bigg(1+\frac{6A}{\sigma}\bigg)+\frac{\rho^2}{2\sigma}+\frac{6B}{\sigma}\bigg],
\end{equation}

\begin{equation} \label{eq17}
-\frac{1}{r^2}=\bigg[\rho\bigg\{1+\bigg(\frac{1+2\omega}{\sigma}\bigg)2A\bigg\}+\frac{3\rho^2}{2\sigma}+\bigg(\frac{1+2\omega}{\sigma}\bigg)2B\bigg],
\end{equation}
\begin{widetext}
\begin{equation} \label{eq18}
-\frac{\lambda^{\prime}\nu^{\prime}}{4}e^{-\lambda}-\frac{e^{-\lambda}\lambda^{\prime}}{2r}=
\bigg[\rho\bigg\{1+\bigg(\frac{1-\omega}{\sigma}\bigg)2A\bigg\}+\frac{3\rho^2}{2\sigma}+\bigg(\frac{1-\omega}{\sigma}\bigg)2B\bigg].
\end{equation}
\end{widetext}
Solving EFE's and using $\rho=\rho_ce^{-\nu(r)}$ with Kuchowicz-like $\nu(r)$ we could obtain analytically second unknown metric potential
\begin{equation}
    e^{-\lambda}=\frac{3 A \rho_c e^{-C r^2}}{C
   D^2 \sigma }-\frac{3 B r^2}{\sigma }+\frac{\rho_c^2 e^{-2 C r^2}}{8 C D^4 \sigma
   }+\frac{\rho_c e^{-C r^2}}{2 C
   D^2}+\log (r)-c_1
\end{equation}
Without the loss of generality it is convenient to assume that $c_1=0$.

\subsection{Exterior region}

The exterior region geometry of the gravastar could be well described by the Schwarzschild metric of the form (with the effective Equation of State $\rho=p=0$):
\begin{widetext}
\begin{equation}
    ds^2 = \bigg(1-\frac{2M}{r}\bigg)dt^2 - \bigg(1-\frac{2M}{r}\bigg)^{-1}dr^2 - r^2 d\theta^2 - r^2 \sin^2 \theta d\phi^2
\end{equation}
\end{widetext}
where $M$ indicates total gravastar mass. For exterior spacetime, since EoS is vacuum one, EFE's has the very simplified form below
\begin{equation}
    e^{-\lambda}\bigg(\frac{\lambda'}{r}-\frac{1}{r^2}\bigg)+\frac{1}{r^2}=\frac{6B}{\sigma}
\end{equation}
Solution is therefore
\begin{equation}
    e^{-\lambda}=1-\frac{2M}{r}-\frac{2B}{\sigma}r^2
\end{equation}
To get rid of effective brane cosmological constant $\Lambda=6B/\sigma$, we will assume that $B=0$ and therefore solution above mimics regular component of Schwarzschild spacetime line element (constant of integration for sake of regularity at the origin is already assumed to vanish).

\section{Physical aspects of gravastars in modified gravity} \label{sec:4}
\subsection{Proper length}

Gravastar shell proper length:
\begin{equation}
    \ell = \int^{R+\epsilon}_{R}\frac{dr}{\sqrt{e^{-\lambda(r)}}}
    \label{eq:23}
\end{equation}
For the Starobinsky, gamma and exponential gravity we have the following proper length of the gravastar shell (using Tolman-Kuchowicz metric potentials):

\begin{figure}[!htbp]
    \centering
    \includegraphics[width=0.7\columnwidth]{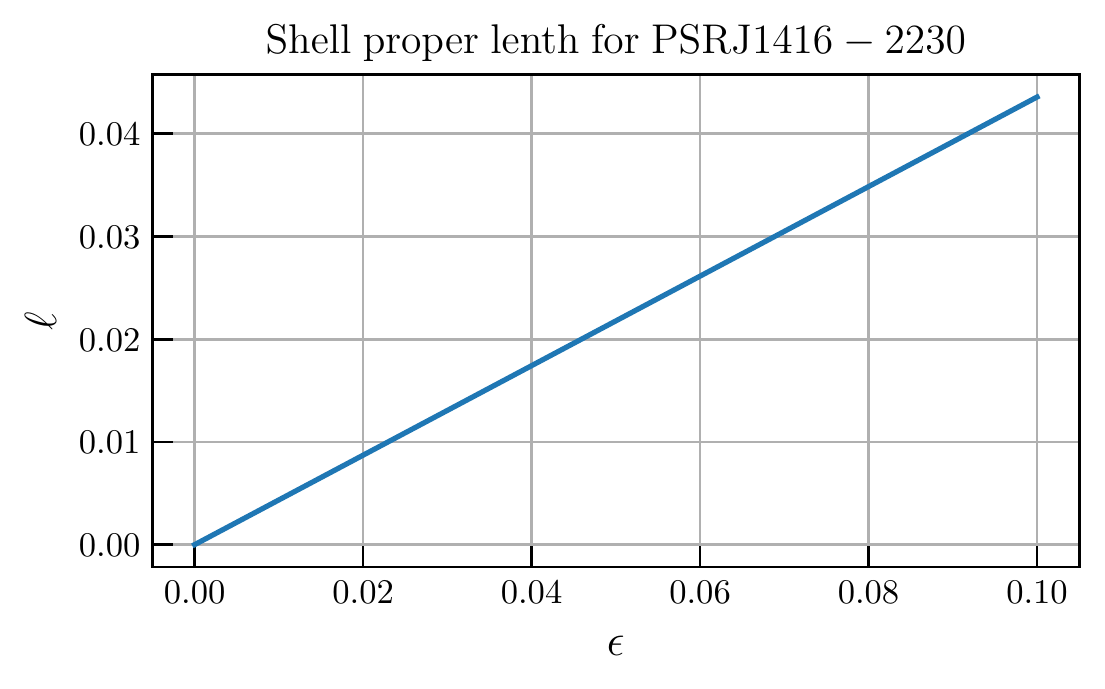}
    \caption{Shell proper length w.r.t shell thickness $\epsilon$. For model we use PSR J1416-2230 compact star with mass $M=1.97M_\odot$ and radius $R=9.69$km \cite{Demorest2010ATN}. To plot the results numerically, we as well assume the same brane tension as for the interior region (namely $\sigma=20.0762$) and the same value for $A$.}
    \label{fig:32}
\end{figure}

On the Figure (\ref{fig:32}) we as usual numerically solved equation (\ref{eq:23}) with $A=1$ and $B=0$ (vanishing effective brane cosmological constant), $R=9.69$ and varying shell thickness. As one may notice, values of the shell proper length $\ell$ grows with $\varepsilon\to\infty$. 
\subsection{Energy}
Energy of the gravastar shell is defined as follows:
\begin{equation}
    \mathcal{E} = \int ^{R+\epsilon}_{R} 4\pi r^2 \rho^{\mathrm{eff}} dr
    \label{eq:24}
\end{equation}
\begin{figure}[!htbp]
    \centering
    \includegraphics[width=0.7\columnwidth]{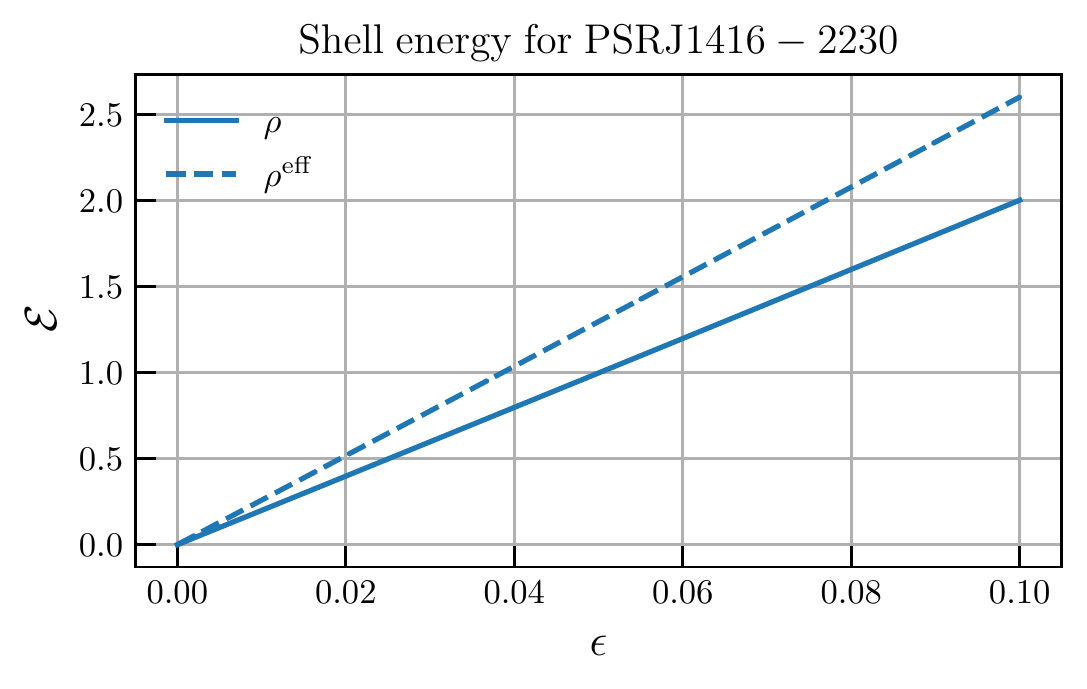}
    \caption{Shell energy w.r.t shell thickness $\epsilon$. For model we use PSR J1416-2230 compact star with mass $M=1.97M_\odot$ and radius $R=9.69$km \cite{Demorest2010ATN}. As usual, we have assumed the same values of brane tension and $A$ constant as in the interior region.}
    \label{fig:332}
\end{figure}

On the Figure (\ref{fig:332}) we numerically solved Equation (\ref{eq:24}) for braneworld gravity with regular energy density $\rho$ and effective energy density $\rho^{\mathrm{eff}}$. Remarkably, as we see if $\epsilon$ grows, $\mathcal{E}\to\infty$, which is expected behavior of shell energy. Also, shell energy for effective energy density is bigger that for regular $\rho$.

\subsection{Entropy}

\begin{figure}[!htbp]
    \centering
    \includegraphics[width=0.7\columnwidth]{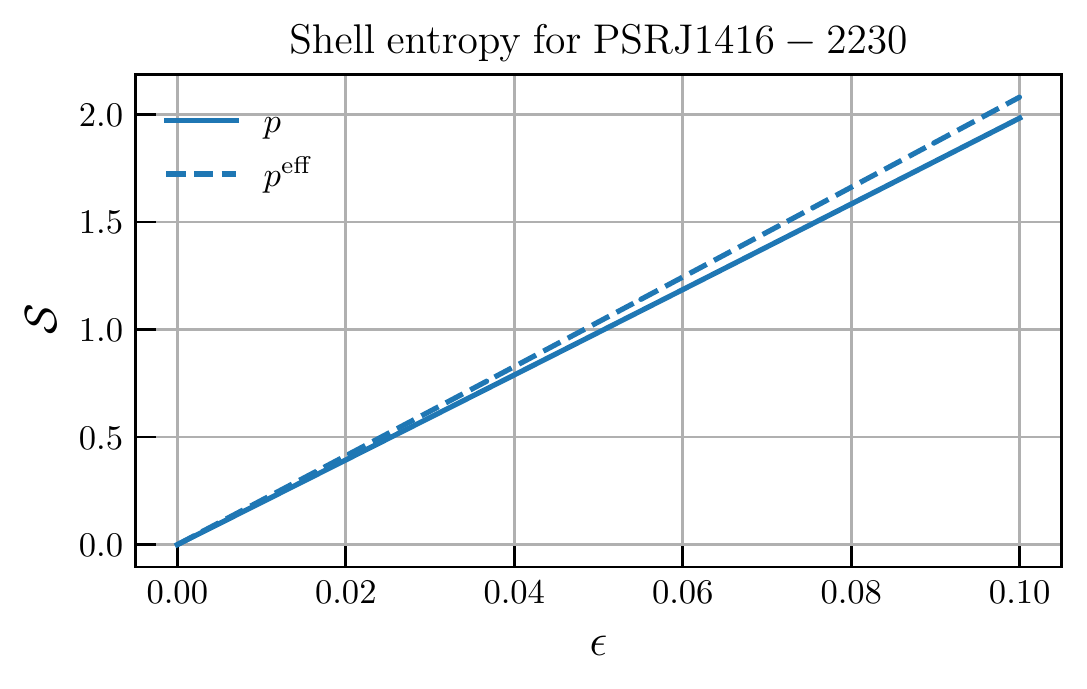}
    \caption{Shell entropy w.r.t shell thickness $\epsilon$. For model we use PSR J1416-2230 compact star with mass $M=1.97M_\odot$ and radius $R=9.69$km \cite{Demorest2010ATN}. As well, we assume that $\rho_c=0.01$, $\xi=0.235$ and $\sigma=20.0762$, $A=1$.}
    \label{fig:111}
\end{figure}

Mazur and Mottola \cite{Mazur2001,Mazur2004} stated that interior region of gravastar must have zero entropy density, which is stable for the single condensate area. But, entropy on the shell is generally non-zero.
The entropy of the relativitstic star system (static) gravastar could be easily determined by the formula below:
\begin{equation}
    S = \int ^{R+\epsilon}_{R} 4\pi r^2 \frac{s(r)}{\sqrt{e^{-\lambda(r)}}}dr
\end{equation}
where 
\begin{equation}
    s(r) = \xi \frac{k_{\mathrm{B}}}{\hbar} \sqrt{\frac{p}{2\pi}}
\end{equation}
We assumed that $k_{\mathrm{B}}=\hbar$.
On the Figure (\ref{fig:111}) variation of entropy within the shell is shown for RS-II braneworld gravastar. During the numerical analysis of the gravastar shell entropy, we noticed that shell entropy grows as shell thickness becomes bigger. Also, for effective pressure (plugged in the definition of $s(r)$ function), entropy is slightly bigger that for regular isotropic pressure $p$.

\subsection{Surface redshift}

\begin{figure}[!htbp]
    \centering
    \includegraphics[width=0.7\columnwidth]{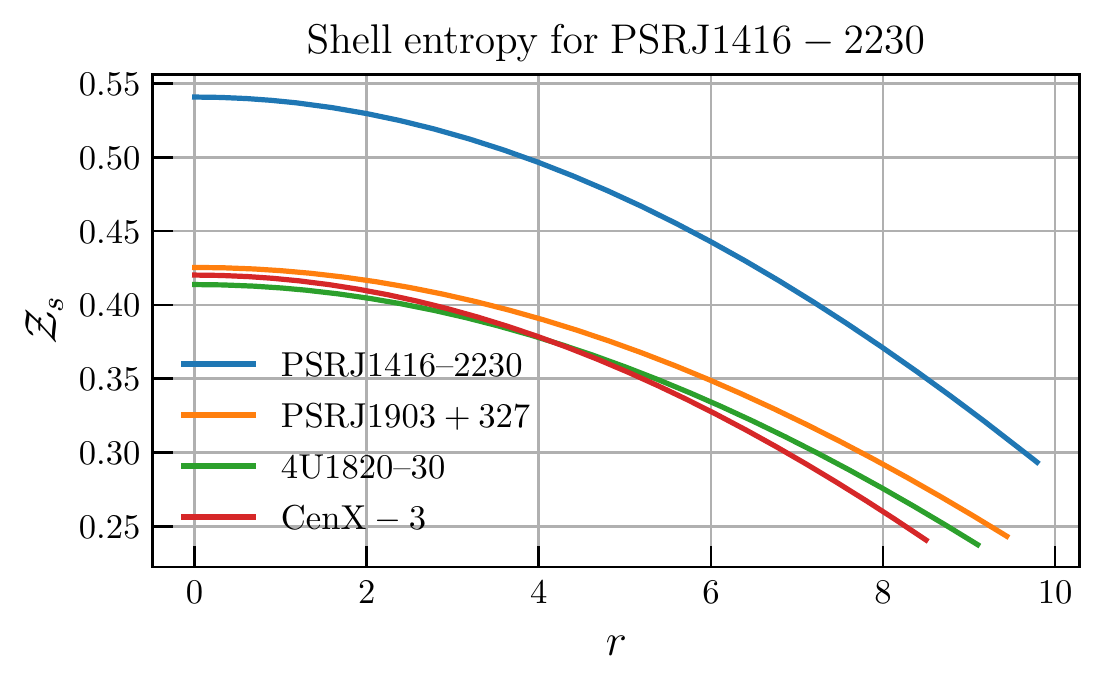}
    \caption{Surface redshift w.r.t radial coordinate $r$.}
    \label{fig:1111}
\end{figure}

Gravastar surface redshift is defined in the following way:
\begin{equation}
    \mathcal{Z}_s = |g_{tt}|^{-1/2} - 1 = \frac{e^{-\frac{1}{2} \Re\bigl(C r^2\bigr)}}{\bigl| D\bigr| }-1
\end{equation}
Surface redshift for the isotropic compact star fluid must not exceed $2$ (for the spacetimes with present cosmological constant surface redshift must not exceed $5$). We plot the surface redshift on the Figure (\ref{fig:1111}) for different compact objects with stellar nature. As we noticed from numerical investigation, for each compact star $\mathcal{Z}_s$ at the whole interior domain does not exceed $2$, which is necessary condition.

\subsection{Adiabatic index}

We could check the dynamical stability of the relativistic stellar against infinitesimal adiabatic perturbations by following the pioneering work of Chandrasekhar \cite{Chandrasekhar1964}. Chandrasekhar predicted that for the relativistic system to be stable the adiabatic index should exceed $4/3$. This adiabatic index is defined as \cite{Maurya2017}:
\begin{equation}
    \Gamma = \frac{p+\rho}{p}\frac{dp}{d\rho}
\end{equation}
Then:
\begin{itemize}
    \item For the interior region with the EoS $p=-\rho$, $\Gamma=0$.
    \item For the intermediate shell region with the EoS $p=\rho$, $\Gamma=2$.
\end{itemize}
Therefore, we could conclude that for gravastars with braneworld gravity formalism, from the adiabatic index interior region is unstable and shell region is stable.

\section{Conclusions} \label{sec:6}

In the present letter we have studied the static and spherically symmetric non-charged gravastars under the framework of braneworld gravity model (we assume that bulk is five dimensional and brane configuration is of RS-II type) with the Kuchowicz metric potential. In this section we want to summarize all of the key results, that was obtained in the paper.

As well, we have derived several physical parameters, such as: interior region mass, proper length, shell energy and entropy, surface redshift and adiabatic index.  We have discussed the nature of this parameters both analytically and graphically. From the numerical solutions we have noted that:
\begin{itemize}
    \item Interior region mass: we investigate the interior region gravitational mass on the Figure (\ref{fig:1}) for both regular and effective energy densities. As one could notice, gravitational mass grows exponentially with radius $R$, which represents the common behavior of interior region DE-like matter.
    \item {Proper length}: The proper length of the gravastar shell $\ell$ is plotted w.r.t. shell thickness. As we noticed, shell proper length is increasing with growing shell thickness. We plotted the results on the Figure (\ref{fig:32}).
    \item Energy: The energy of the shell $\mathcal{E}$  was probed and illustrated on the Figure (\ref{fig:332}). As well, the shell energy behaves as expected.
    \item Entropy: For the interior region, entropy density $S$ is zero, but for the shell region generally not. We plotted the entropy for regular braneworld gravity on Figure (\ref{fig:111}). As we see, on-shell entropy monotonously grow with shell thickness.
    \item Surface redshift: from the values of the surface redshift $\mathcal{Z}_s$ we could judge whether the compact object is stable or not. For the isotropic fluid (with $p_r=p_t$), surface redshift must not exceed the value of $2$, which is obeyed for some compact stars and results are plotted on the Figure (\ref{fig:1111}).
\end{itemize}

As we already said, gravastars are  usually separated on three different regions: interior, shell and exterior. With Kuchowicz metric potential, from Equation of State for each region we have derived analytically second unknown metric potential for $g_{rr}$ component.

In the end, we came to the conclusion that we have derived new, non-singular and horizonless gravastar model for braneworld gravity with the impact of Kuchowicz metric potential. Generally, with special metric potential it is more challenging to obtain physically acceptable solutions, but, as one could notice, interest to the Tolman-Kuchowicz metric potentials for the compact objects (exotic stars for example) have grown in this decade (see \cite{Rej2021,Shamir2020b,Biswas2020,Majid2020,Jasim2018,FarasatShamir2020}), and thus it is important to test the Kuchowicz spacetime on static gravastars.

\section*{Acknowledgment}
PKS acknowledges National Board for Higher Mathematics (NBHM), No.: 02011/3/2022 NBHM(R.P.)/R\&D II/2152 Dt.14.02.2022, Govt. of India under Department of Atomic Energy (DAE). We are thankful to the honorable anonymous referee for helpful comments, which have significantly improved our work in terms of research quality and presentation.


\begin{thebibliography}{42}%
\makeatletter
\providecommand \@ifxundefined [1]{%
 \@ifx{#1\undefined}
}%
\providecommand \@ifnum [1]{%
 \ifnum #1\expandafter \@firstoftwo
 \else \expandafter \@secondoftwo
 \fi
}%
\providecommand \@ifx [1]{%
 \ifx #1\expandafter \@firstoftwo
 \else \expandafter \@secondoftwo
 \fi
}%
\providecommand \natexlab [1]{#1}%
\providecommand \enquote  [1]{``#1''}%
\providecommand \bibnamefont  [1]{#1}%
\providecommand \bibfnamefont [1]{#1}%
\providecommand \citenamefont [1]{#1}%
\providecommand \href@noop [0]{\@secondoftwo}%
\providecommand \href [0]{\begingroup \@sanitize@url \@href}%
\providecommand \@href[1]{\@@startlink{#1}\@@href}%
\providecommand \@@href[1]{\endgroup#1\@@endlink}%
\providecommand \@sanitize@url [0]{\catcode `\\12\catcode `\$12\catcode
  `\&12\catcode `\#12\catcode `\^12\catcode `\_12\catcode `\%12\relax}%
\providecommand \@@startlink[1]{}%
\providecommand \@@endlink[0]{}%
\providecommand \url  [0]{\begingroup\@sanitize@url \@url }%
\providecommand \@url [1]{\endgroup\@href {#1}{\urlprefix }}%
\providecommand \urlprefix  [0]{URL }%
\providecommand \Eprint [0]{\href }%
\providecommand \doibase [0]{https://doi.org/}%
\providecommand \selectlanguage [0]{\@gobble}%
\providecommand \bibinfo  [0]{\@secondoftwo}%
\providecommand \bibfield  [0]{\@secondoftwo}%
\providecommand \translation [1]{[#1]}%
\providecommand \BibitemOpen [0]{}%
\providecommand \bibitemStop [0]{}%
\providecommand \bibitemNoStop [0]{.\EOS\space}%
\providecommand \EOS [0]{\spacefactor3000\relax}%
\providecommand \BibitemShut  [1]{\csname bibitem#1\endcsname}%
\let\auto@bib@innerbib\@empty
%</preamble>
\bibitem [{\citenamefont {Pani}\ \emph {et~al.}(2010)\citenamefont {Pani},
  \citenamefont {Berti}, \citenamefont {Cardoso}, \citenamefont {Chen},\ and\
  \citenamefont {Norte}}]{Pani2010}%
  \BibitemOpen
  \bibfield  {author} {\bibinfo {author} {\bibfnamefont {P.}~\bibnamefont
  {Pani}}, \bibinfo {author} {\bibfnamefont {E.}~\bibnamefont {Berti}},
  \bibinfo {author} {\bibfnamefont {V.}~\bibnamefont {Cardoso}}, \bibinfo
  {author} {\bibfnamefont {Y.}~\bibnamefont {Chen}},\ and\ \bibinfo {author}
  {\bibfnamefont {R.}~\bibnamefont {Norte}},\ }\bibfield  {title} {\bibinfo
  {title} {Gravitational-wave signature of a thin-shell gravastar},\ }\href
  {https://doi.org/10.1088/1742-6596/222/1/012032} {\bibfield  {journal}
  {\bibinfo  {journal} {Journal of Physics: Conference Series}\ }\textbf
  {\bibinfo {volume} {222}},\ \bibinfo {pages} {012032} (\bibinfo {year}
  {2010})}\BibitemShut {NoStop}%
\bibitem [{\citenamefont {Chan}\ \emph {et~al.}(2010)\citenamefont {Chan},
  \citenamefont {da~Silva}, \citenamefont {Rocha},\ and\ \citenamefont
  {Wang}}]{Chan2010}%
  \BibitemOpen
  \bibfield  {author} {\bibinfo {author} {\bibfnamefont {R.}~\bibnamefont
  {Chan}}, \bibinfo {author} {\bibfnamefont {M.~F.~A.}\ \bibnamefont
  {da~Silva}}, \bibinfo {author} {\bibfnamefont {P.}~\bibnamefont {Rocha}},\
  and\ \bibinfo {author} {\bibfnamefont {A.}~\bibnamefont {Wang}},\ }\bibfield
  {title} {\bibinfo {title} {Gravastars or black holes as consequence of
  Einstein’s theory of gravity},\ }\href {https://doi.org/10.1063/1.3462687}
  {\bibfield  {journal} {\bibinfo  {journal} {AIP Conference Proceedings}\
  }\textbf {\bibinfo {volume} {1241}},\ \bibinfo {pages} {571} (\bibinfo {year}
  {2010})}\BibitemShut {NoStop}%
\bibitem [{\citenamefont {Kubo}\ and\ \citenamefont {Sakai}(2016)}]{Kubo2016}%
  \BibitemOpen
  \bibfield  {author} {\bibinfo {author} {\bibfnamefont {T.}~\bibnamefont
  {Kubo}}\ and\ \bibinfo {author} {\bibfnamefont {N.}~\bibnamefont {Sakai}},\
  }\bibfield  {title} {\bibinfo {title} {Gravitational lensing by gravastars},\
  }\href {https://doi.org/10.1103/PhysRevD.93.084051} {\bibfield  {journal}
  {\bibinfo  {journal} {Phys. Rev. D}\ }\textbf {\bibinfo {volume} {93}},\
  \bibinfo {pages} {084051} (\bibinfo {year} {2016})}\BibitemShut {NoStop}%
\bibitem [{\citenamefont {Banerjee}\ \emph {et~al.}(2020)\citenamefont
  {Banerjee}, \citenamefont {Ghosh}, \citenamefont {Paul},\ and\ \citenamefont
  {Rahaman}}]{Banerjee2020}%
  \BibitemOpen
  \bibfield  {author} {\bibinfo {author} {\bibfnamefont {S.}~\bibnamefont
  {Banerjee}}, \bibinfo {author} {\bibfnamefont {S.}~\bibnamefont {Ghosh}},
  \bibinfo {author} {\bibfnamefont {N.}~\bibnamefont {Paul}},\ and\ \bibinfo
  {author} {\bibfnamefont {F.}~\bibnamefont {Rahaman}},\ }\bibfield  {title}
  {\bibinfo {title} {Study of gravastars in finslerian geometry},\ }\href
  {https://doi.org/10.1140/epjp/s13360-020-00230-0} {\bibfield  {journal}
  {\bibinfo  {journal} {The European Physical Journal Plus}\ }\textbf {\bibinfo
  {volume} {135}},\ \bibinfo {pages} {185} (\bibinfo {year}
  {2020})}\BibitemShut {NoStop}%
\bibitem [{\citenamefont {Ghosh}\ \emph {et~al.}(2020)\citenamefont {Ghosh},
  \citenamefont {Kanfon}, \citenamefont {Das}, \citenamefont {Houndjo},
  \citenamefont {Salako},\ and\ \citenamefont {Ray}}]{Ghosh2020}%
  \BibitemOpen
  \bibfield  {author} {\bibinfo {author} {\bibfnamefont {S.}~\bibnamefont
  {Ghosh}}, \bibinfo {author} {\bibfnamefont {A.~D.}\ \bibnamefont {Kanfon}},
  \bibinfo {author} {\bibfnamefont {A.}~\bibnamefont {Das}}, \bibinfo {author}
  {\bibfnamefont {M.~J.~S.}\ \bibnamefont {Houndjo}}, \bibinfo {author}
  {\bibfnamefont {I.~G.}\ \bibnamefont {Salako}},\ and\ \bibinfo {author}
  {\bibfnamefont {S.}~\bibnamefont {Ray}},\ }\bibfield  {title} {\bibinfo
  {title} {{Gravastars in $f(\mathbb{T},\mathcal{T})$ gravity}},\ }\href
  {https://doi.org/10.1142/S0217751X20500177} {\bibfield  {journal} {\bibinfo
  {journal} {Int. J. Mod. Phys. A}\ }\textbf {\bibinfo {volume} {35}},\
  \bibinfo {pages} {2050017} (\bibinfo {year} {2020})}\BibitemShut {NoStop}%
\bibitem [{\citenamefont {Shamir}\ and\ \citenamefont
  {Fayyaz}(2020)}]{Shamir2020}%
  \BibitemOpen
  \bibfield  {author} {\bibinfo {author} {\bibfnamefont {M.~F.}\ \bibnamefont
  {Shamir}}\ and\ \bibinfo {author} {\bibfnamefont {I.}~\bibnamefont
  {Fayyaz}},\ }\bibfield  {title} {\bibinfo {title} {Traversable wormhole
  solutions in $f(\mathcal{R})$ gravity via karmarkar condition},\ }\href
  {https://doi.org/10.1140/epjc/s10052-020-08689-y} {\bibfield  {journal}
  {\bibinfo  {journal} {The European Physical Journal C}\ }\textbf {\bibinfo
  {volume} {80}},\ \bibinfo {pages} {1102} (\bibinfo {year}
  {2020})}\BibitemShut {NoStop}%
\bibitem [{\citenamefont {Abbas}\ and\ \citenamefont
  {Majeed}(2020)}]{Abbas2020}%
  \BibitemOpen
  \bibfield  {author} {\bibinfo {author} {\bibfnamefont {G.}~\bibnamefont
  {Abbas}}\ and\ \bibinfo {author} {\bibfnamefont {K.}~\bibnamefont {Majeed}},\
  }\bibfield  {title} {\bibinfo {title} {Isotropic gravastar model in Rastall
  gravity},\ }\href {https://doi.org/10.1155/2020/8861168} {\bibfield
  {journal} {\bibinfo  {journal} {Advances in Astronomy}\ }\textbf {\bibinfo
  {volume} {2020}},\ \bibinfo {pages} {8861168} (\bibinfo {year}
  {2020})}\BibitemShut {NoStop}%
\bibitem [{\citenamefont {Kuhfittig}\ and\ \citenamefont
  {Gladney}(2020)}]{Kuhfittig2020}%
  \BibitemOpen
  \bibfield  {author} {\bibinfo {author} {\bibfnamefont {P.~K.~F.}\
  \bibnamefont {Kuhfittig}}\ and\ \bibinfo {author} {\bibfnamefont {V.~D.}\
  \bibnamefont {Gladney}},\ }\bibfield  {title} {\bibinfo {title} {Seeking
  connections between wormholes, gravastars, and black holes via noncommutative
  geometry},\ }\href {https://doi.org/10.1142/S0217732320500595} {\bibfield
  {journal} {\bibinfo  {journal} {Modern Physics Letters A}\ }\textbf {\bibinfo
  {volume} {35}},\ \bibinfo {pages} {2050059} (\bibinfo {year}
  {2020})}\BibitemShut {NoStop}%
\bibitem [{\citenamefont {{Mottola}}\ and\ \citenamefont
  {{Mazur}}(2002)}]{Mazur2001}%
  \BibitemOpen
  \bibfield  {author} {\bibinfo {author} {\bibfnamefont {E.}~\bibnamefont
  {{Mottola}}}\ and\ \bibinfo {author} {\bibfnamefont {P.~O.}\ \bibnamefont
  {{Mazur}}},\ }\bibfield  {title} {\bibinfo {title} {{Gravitational Condensate
  Stars: An Alternative to Black Holes}},\ }in\ \href@noop {} {\emph {\bibinfo
  {booktitle} {APS April Meeting Abstracts}}},\ \bibinfo {series and number}
  {APS Meeting Abstracts}\ (\bibinfo {year} {2002})\ p.\ \bibinfo {pages}
  {I12.011}\BibitemShut {NoStop}%
\bibitem [{\citenamefont {Mazur}\ and\ \citenamefont
  {Mottola}(2004)}]{Mazur2004}%
  \BibitemOpen
  \bibfield  {author} {\bibinfo {author} {\bibfnamefont {P.}~\bibnamefont
  {Mazur}}\ and\ \bibinfo {author} {\bibfnamefont {E.}~\bibnamefont
  {Mottola}},\ }\bibfield  {title} {\bibinfo {title} {Gravitational vacuum
  condensate stars},\ }\href {https://doi.org/10.1073/pnas.0402717101}
  {\bibfield  {journal} {\bibinfo  {journal} {Proceedings of the National
  Academy of Sciences of the United States of America}\ }\textbf {\bibinfo
  {volume} {101}},\ \bibinfo {pages} {9545} (\bibinfo {year}
  {2004})}\BibitemShut {NoStop}%
\bibitem [{\citenamefont {Buchdahl}(1970{\natexlab{a}})}]{Buchdahl1970}%
  \BibitemOpen
  \bibfield  {author} {\bibinfo {author} {\bibfnamefont {H.~A.}\ \bibnamefont
  {Buchdahl}},\ }\bibfield  {title} {\bibinfo {title} {{Non-Linear Lagrangians
  and Cosmological Theory}},\ }\href {https://doi.org/10.1093/mnras/150.1.1}
  {\bibfield  {journal} {\bibinfo  {journal} {Monthly Notices of the Royal
  Astronomical Society}\ }\textbf {\bibinfo {volume} {150}},\ \bibinfo {pages}
  {1} (\bibinfo {year} {1970}{\natexlab{a}})}\BibitemShut {NoStop}%
\bibitem [{\citenamefont {Brooker}\ \emph {et~al.}(2016)\citenamefont
  {Brooker}, \citenamefont {Odintsov},\ and\ \citenamefont
  {Woodard}}]{Brooker2016}%
  \BibitemOpen
  \bibfield  {author} {\bibinfo {author} {\bibfnamefont {D.}~\bibnamefont
  {Brooker}}, \bibinfo {author} {\bibfnamefont {S.}~\bibnamefont {Odintsov}},\
  and\ \bibinfo {author} {\bibfnamefont {R.}~\bibnamefont {Woodard}},\
  }\bibfield  {title} {\bibinfo {title} {Precision predictions for the
  primordial power spectra from $f(\mathcal{R})$ models of inflation},\ }\href
  {https://doi.org/10.1016/j.nuclphysb.2016.08.010} {\bibfield  {journal}
  {\bibinfo  {journal} {Nuclear Physics B}\ }\textbf {\bibinfo {volume}
  {911}},\ \bibinfo {pages} {318} (\bibinfo {year} {2016})}\BibitemShut
  {NoStop}%
\bibitem [{\citenamefont {Huang}(2014)}]{Huang2013}%
  \BibitemOpen
  \bibfield  {author} {\bibinfo {author} {\bibfnamefont {Q.-G.}\ \bibnamefont
  {Huang}},\ }\bibfield  {title} {\bibinfo {title} {A polynomial
  $f(\mathcal{R})$ inflation model},\ }\href
  {https://doi.org/10.1088/1475-7516/2014/02/035} {\bibfield  {journal}
  {\bibinfo  {journal} {Journal of Cosmology and Astroparticle Physics}\
  }\textbf {\bibinfo {volume} {2014}}\bibinfo  {number} { (02)},\ \bibinfo
  {pages} {035}}\BibitemShut {NoStop}%
\bibitem [{\citenamefont {Starobinsky}(1980)}]{Starobinsky1980}%
  \BibitemOpen
\bibfield  {number} {  }\bibfield  {author} {\bibinfo {author} {\bibfnamefont
  {A.}~\bibnamefont {Starobinsky}},\ }\bibfield  {title} {\bibinfo {title} {A
  new type of isotropic cosmological models without singularity},\ }\href
  {https://doi.org/https://doi.org/10.1016/0370-2693(80)90670-X} {\bibfield
  {journal} {\bibinfo  {journal} {Physics Letters B}\ }\textbf {\bibinfo
  {volume} {91}},\ \bibinfo {pages} {99} (\bibinfo {year} {1980})}\BibitemShut
  {NoStop}%
\bibitem [{\citenamefont {Capozziello}\ and\ \citenamefont {{De
  Laurentis}}(2011)}]{Capozziello2011}%
  \BibitemOpen
  \bibfield  {author} {\bibinfo {author} {\bibfnamefont {S.}~\bibnamefont
  {Capozziello}}\ and\ \bibinfo {author} {\bibfnamefont {M.}~\bibnamefont {{De
  Laurentis}}},\ }\bibfield  {title} {\bibinfo {title} {Extended theories of
  gravity},\ }\href
  {https://doi.org/https://doi.org/10.1016/j.physrep.2011.09.003} {\bibfield
  {journal} {\bibinfo  {journal} {Physics Reports}\ }\textbf {\bibinfo {volume}
  {509}},\ \bibinfo {pages} {167} (\bibinfo {year} {2011})}\BibitemShut
  {NoStop}%
\bibitem [{\citenamefont {Nojiri}\ \emph {et~al.}(2017)\citenamefont {Nojiri},
  \citenamefont {Odintsov},\ and\ \citenamefont {Oikonomou}}]{Nojiri2017}%
  \BibitemOpen
  \bibfield  {author} {\bibinfo {author} {\bibfnamefont {S.}~\bibnamefont
  {Nojiri}}, \bibinfo {author} {\bibfnamefont {S.}~\bibnamefont {Odintsov}},\
  and\ \bibinfo {author} {\bibfnamefont {V.}~\bibnamefont {Oikonomou}},\
  }\bibfield  {title} {\bibinfo {title} {Modified gravity theories on a
  nutshell: Inflation, bounce and late-time evolution},\ }\href
  {https://doi.org/https://doi.org/10.1016/j.physrep.2017.06.001} {\bibfield
  {journal} {\bibinfo  {journal} {Physics Reports}\ }\textbf {\bibinfo {volume}
  {692}},\ \bibinfo {pages} {1} (\bibinfo {year} {2017})}\BibitemShut {NoStop}%
\bibitem [{\citenamefont {Das}\ \emph {et~al.}(2017)\citenamefont {Das},
  \citenamefont {Ghosh}, \citenamefont {Guha}, \citenamefont {Das},
  \citenamefont {Rahaman},\ and\ \citenamefont {Ray}}]{Das2017}%
  \BibitemOpen
  \bibfield  {author} {\bibinfo {author} {\bibfnamefont {A.}~\bibnamefont
  {Das}}, \bibinfo {author} {\bibfnamefont {S.}~\bibnamefont {Ghosh}}, \bibinfo
  {author} {\bibfnamefont {B.~K.}\ \bibnamefont {Guha}}, \bibinfo {author}
  {\bibfnamefont {S.}~\bibnamefont {Das}}, \bibinfo {author} {\bibfnamefont
  {F.}~\bibnamefont {Rahaman}},\ and\ \bibinfo {author} {\bibfnamefont
  {S.}~\bibnamefont {Ray}},\ }\bibfield  {title} {\bibinfo {title} {{Gravastars
  in $f(R,T)$ gravity}},\ }\href {https://doi.org/10.1103/PhysRevD.95.124011}
  {\bibfield  {journal} {\bibinfo  {journal} {Phys. Rev. D}\ }\textbf {\bibinfo
  {volume} {95}},\ \bibinfo {pages} {124011} (\bibinfo {year}
  {2017})}\BibitemShut {NoStop}%
\bibitem [{\citenamefont {Debnath}(2019)}]{Debnath2019}%
  \BibitemOpen
  \bibfield  {author} {\bibinfo {author} {\bibfnamefont {U.}~\bibnamefont
  {Debnath}},\ }\bibfield  {title} {\bibinfo {title} {Charge gravastars in
  $f(\mathcal{T})$ modified gravity},\ }\href
  {https://doi.org/10.1140/epjc/s10052-019-7013-z} {\bibfield  {journal}
  {\bibinfo  {journal} {The European Physical Journal C}\ }\textbf {\bibinfo
  {volume} {79}},\ \bibinfo {pages} {499} (\bibinfo {year} {2019})}\BibitemShut
  {NoStop}%
\bibitem [{\citenamefont {Randall}\ and\ \citenamefont
  {Sundrum}(1999)}]{Randall1999}%
  \BibitemOpen
  \bibfield  {author} {\bibinfo {author} {\bibfnamefont {L.}~\bibnamefont
  {Randall}}\ and\ \bibinfo {author} {\bibfnamefont {R.}~\bibnamefont
  {Sundrum}},\ }\bibfield  {title} {\bibinfo {title} {Large mass hierarchy from
  a small extra dimension},\ }\href
  {https://doi.org/10.1103/PhysRevLett.83.3370} {\bibfield  {journal} {\bibinfo
   {journal} {Phys. Rev. Lett.}\ }\textbf {\bibinfo {volume} {83}},\ \bibinfo
  {pages} {3370} (\bibinfo {year} {1999})}\BibitemShut {NoStop}%
  \bibitem [{\citenamefont {Randall}\ and\ \citenamefont
  {Sundrum}(1999)}]{PhysRevLett.83.4690}%
  \BibitemOpen
  \bibfield  {author} {\bibinfo {author} {\bibfnamefont {L.}~\bibnamefont
  {Randall}}\ and\ \bibinfo {author} {\bibfnamefont {R.}~\bibnamefont
  {Sundrum}},\ }\bibfield  {title} {\bibinfo {title} {An alternative to
  compactification},\ }\href {https://doi.org/10.1103/PhysRevLett.83.4690}
  {\bibfield  {journal} {\bibinfo  {journal} {Phys. Rev. Lett.}\ }\textbf
  {\bibinfo {volume} {83}},\ \bibinfo {pages} {4690} (\bibinfo {year}
  {1999})}\BibitemShut {NoStop}%
\bibitem [{\citenamefont {Buchdahl}(1970{\natexlab{b}})}]{Buchdahl:1983zz}%
  \BibitemOpen
  \bibfield  {author} {\bibinfo {author} {\bibfnamefont {H.~A.}\ \bibnamefont
  {Buchdahl}},\ }\bibfield  {title} {\bibinfo {title} {{Non-linear Lagrangians
  and cosmological theory}},\ }\href@noop {} {\bibfield  {journal} {\bibinfo
  {journal} {Mon. Not. Roy. Astron. Soc.}\ }\textbf {\bibinfo {volume} {150}},\
  \bibinfo {pages} {1} (\bibinfo {year} {1970}{\natexlab{b}})}\BibitemShut
  {NoStop}%
\bibitem [{\citenamefont {Sotiriou}\ and\ \citenamefont
  {Faraoni}(2010)}]{Sotiriou2010}%
  \BibitemOpen
  \bibfield  {author} {\bibinfo {author} {\bibfnamefont {T.}~\bibnamefont
  {Sotiriou}}\ and\ \bibinfo {author} {\bibfnamefont {V.}~\bibnamefont
  {Faraoni}},\ }\bibfield  {title} {\bibinfo {title} {$f(\mathcal{R})$ theories
  of gravity},\ }\href {https://doi.org/10.1103/RevModPhys.82.451} {\bibfield
  {journal} {\bibinfo  {journal} {REVIEWS OF MODERN PHYSICS}\ }\textbf
  {\bibinfo {volume} {82}},\ \bibinfo {pages} {451} (\bibinfo {year}
  {2010})}\BibitemShut {NoStop}%
\bibitem [{\citenamefont {Sengupta}\ \emph {et~al.}(2020)\citenamefont
  {Sengupta}, \citenamefont {Ghosh}, \citenamefont {Ray}, \citenamefont
  {Mishra},\ and\ \citenamefont {Tripathy}}]{Sengupta2020}%
  \BibitemOpen
  \bibfield  {author} {\bibinfo {author} {\bibfnamefont {R.}~\bibnamefont
  {Sengupta}}, \bibinfo {author} {\bibfnamefont {S.}~\bibnamefont {Ghosh}},
  \bibinfo {author} {\bibfnamefont {S.}~\bibnamefont {Ray}}, \bibinfo {author}
  {\bibfnamefont {B.}~\bibnamefont {Mishra}},\ and\ \bibinfo {author}
  {\bibfnamefont {S.}~\bibnamefont {Tripathy}},\ }\bibfield  {title} {\bibinfo
  {title} {Gravastar in the framework of braneworld gravity},\ }\href
  {https://doi.org/10.1103/PhysRevD.102.024037} {\bibfield  {journal} {\bibinfo
   {journal} {Physical review D: Particles and fields}\ }\textbf {\bibinfo
  {volume} {102}},\ \bibinfo {pages} {024037} (\bibinfo {year}
  {2020})}\BibitemShut {NoStop}%
\bibitem [{\citenamefont {Arba\~nil}\ \emph {et~al.}(2019)\citenamefont
  {Arba\~nil}, \citenamefont {Moraes},\ and\ \citenamefont
  {Malheiro}}]{Arbanil:2019xfi}%
  \BibitemOpen
  \bibfield  {author} {\bibinfo {author} {\bibfnamefont {J.~D.~V.}\
  \bibnamefont {Arba\~nil}}, \bibinfo {author} {\bibfnamefont {P.~H. R.~S.}\
  \bibnamefont {Moraes}},\ and\ \bibinfo {author} {\bibfnamefont
  {M.}~\bibnamefont {Malheiro}},\ }\bibfield  {title} {\bibinfo {title}
  {{Gravastar model in Randall-Sundrum braneworld}},\ }\href
  {https://doi.org/10.1088/1361-6382/ab47d6} {\bibfield  {journal} {\bibinfo
  {journal} {Class. Quant. Grav.}\ }\textbf {\bibinfo {volume} {36}},\ \bibinfo
  {pages} {235012} (\bibinfo {year} {2019})}\BibitemShut {NoStop}%
    \bibitem [{\citenamefont {Banerjee}\ \emph {et~al.}(2016)\citenamefont
  {Banerjee}, \citenamefont {Rahaman}, \citenamefont {Islam},\ and\
  \citenamefont {Govender}}]{Banerjee:2015ipa}%
  \BibitemOpen
  \bibfield  {author} {\bibinfo {author} {\bibfnamefont {A.}~\bibnamefont
  {Banerjee}}, \bibinfo {author} {\bibfnamefont {F.}~\bibnamefont {Rahaman}},
  \bibinfo {author} {\bibfnamefont {S.}~\bibnamefont {Islam}},\ and\ \bibinfo
  {author} {\bibfnamefont {M.}~\bibnamefont {Govender}},\ }\bibfield  {title}
  {\bibinfo {title} {{Braneworld gravastars admitting conformal motion}},\
  }\href {https://doi.org/10.1140/epjc/s10052-016-3887-1} {\bibfield  {journal}
  {\bibinfo  {journal} {Eur. Phys. J. C}\ }\textbf {\bibinfo {volume} {76}},\
  \bibinfo {pages} {34} (\bibinfo {year} {2016})},\ \Eprint
  {https://arxiv.org/abs/1510.05939} {arXiv:1510.05939 [gr-qc]} \BibitemShut
  {NoStop}%
\bibitem [{\citenamefont {Maartens}(2004)}]{Maartens:2003tw}%
  \BibitemOpen
  \bibfield  {author} {\bibinfo {author} {\bibfnamefont {R.}~\bibnamefont
  {Maartens}},\ }\bibfield  {title} {\bibinfo {title} {{Brane world gravity}},\
  }\href {https://doi.org/10.12942/lrr-2004-7} {\bibfield  {journal} {\bibinfo
  {journal} {Living Rev. Rel.}\ }\textbf {\bibinfo {volume} {7}},\ \bibinfo
  {pages} {7} (\bibinfo {year} {2004})},\ \Eprint
  {https://arxiv.org/abs/gr-qc/0312059} {arXiv:gr-qc/0312059} \BibitemShut
  {NoStop}%
  \bibitem [{\citenamefont {Shiromizu}\ \emph {et~al.}(2000)\citenamefont
  {Shiromizu}, \citenamefont {Maeda},\ and\ \citenamefont
  {Sasaki}}]{PhysRevD.62.024012}%
  \BibitemOpen
  \bibfield  {author} {\bibinfo {author} {\bibfnamefont {T.}~\bibnamefont
  {Shiromizu}}, \bibinfo {author} {\bibfnamefont {K.-i.}\ \bibnamefont
  {Maeda}},\ and\ \bibinfo {author} {\bibfnamefont {M.}~\bibnamefont
  {Sasaki}},\ }\bibfield  {title} {\bibinfo {title} {The Einstein equations on
  the 3-brane world},\ }\href {https://doi.org/10.1103/PhysRevD.62.024012}
  {\bibfield  {journal} {\bibinfo  {journal} {Phys. Rev. D}\ }\textbf {\bibinfo
  {volume} {62}},\ \bibinfo {pages} {024012} (\bibinfo {year}
  {2000})}\BibitemShut {NoStop}%
\bibitem [{\citenamefont {Oppenheimer}\ and\ \citenamefont
  {Volkoff}(1939)}]{Oppenheimer1939}%
  \BibitemOpen
  \bibfield  {author} {\bibinfo {author} {\bibfnamefont {J.~R.}\ \bibnamefont
  {Oppenheimer}}\ and\ \bibinfo {author} {\bibfnamefont {G.~M.}\ \bibnamefont
  {Volkoff}},\ }\bibfield  {title} {\bibinfo {title} {On massive neutron
  cores},\ }\href {https://doi.org/10.1103/PhysRev.55.374} {\bibfield
  {journal} {\bibinfo  {journal} {Phys. Rev.}\ }\textbf {\bibinfo {volume}
  {55}},\ \bibinfo {pages} {374} (\bibinfo {year} {1939})}\BibitemShut
  {NoStop}%
\bibitem [{\citenamefont {Ponce~de Leon}(1993)}]{Poncede1993}%
  \BibitemOpen
  \bibfield  {author} {\bibinfo {author} {\bibfnamefont {J.}~\bibnamefont
  {Ponce~de Leon}},\ }\bibfield  {title} {\bibinfo {title} {Limiting
  configurations allowed by the energy conditions},\ }\href
  {https://doi.org/10.1007/BF00763756} {\bibfield  {journal} {\bibinfo
  {journal} {General Relativity and Gravitation}\ }\textbf {\bibinfo {volume}
  {25}},\ \bibinfo {pages} {1123} (\bibinfo {year} {1993})}\BibitemShut
  {NoStop}%
\bibitem [{\citenamefont {Rahaman}\ \emph {et~al.}(2014)\citenamefont
  {Rahaman}, \citenamefont {Kuhfittig}, \citenamefont {Ray},\ and\
  \citenamefont {Islam}}]{Rahaman2014}%
  \BibitemOpen
  \bibfield  {author} {\bibinfo {author} {\bibfnamefont {F.}~\bibnamefont
  {Rahaman}}, \bibinfo {author} {\bibfnamefont {P.~K.~F.}\ \bibnamefont
  {Kuhfittig}}, \bibinfo {author} {\bibfnamefont {S.}~\bibnamefont {Ray}},\
  and\ \bibinfo {author} {\bibfnamefont {N.}~\bibnamefont {Islam}},\ }\bibfield
   {title} {\bibinfo {title} {Possible existence of wormholes in the galactic
  halo region},\ }\href {https://doi.org/10.1140/epjc/s10052-014-2750-5}
  {\bibfield  {journal} {\bibinfo  {journal} {The European Physical Journal C}\
  }\textbf {\bibinfo {volume} {74}},\ \bibinfo {pages} {2750} (\bibinfo {year}
  {2014})}\BibitemShut {NoStop}%
\bibitem [{\citenamefont {Tolman}(1939)}]{Tolman1939}%
  \BibitemOpen
  \bibfield  {author} {\bibinfo {author} {\bibfnamefont {R.~C.}\ \bibnamefont
  {Tolman}},\ }\bibfield  {title} {\bibinfo {title} {Static solutions of
  Einstein's field equations for spheres of fluid},\ }\href
  {https://doi.org/10.1103/PhysRev.55.364} {\bibfield  {journal} {\bibinfo
  {journal} {Phys. Rev.}\ }\textbf {\bibinfo {volume} {55}},\ \bibinfo {pages}
  {364} (\bibinfo {year} {1939})}\BibitemShut {NoStop}%
\bibitem [{\citenamefont {Kuchowicz}(1968)}]{Kuchowicz1968}%
  \BibitemOpen
  \bibfield  {author} {\bibinfo {author} {\bibfnamefont {B.}~\bibnamefont
  {Kuchowicz}},\ }\bibfield  {title} {\bibinfo {title} {General relativistic
  fluid spheres. i. new solutions for spherically symmetric matter
  distributions.},\ }\bibfield  {journal} {\bibinfo  {journal} {Acta Phys.
  Pol., \textbf{33}: 541-63(Apr. 1968).}\ }\href@noop {} {} (\bibinfo {year}
  {1968})\BibitemShut {NoStop}%
\bibitem [{\citenamefont {Bhar}(2021)}]{BHAR2021100879}%
  \BibitemOpen
  \bibfield  {author} {\bibinfo {author} {\bibfnamefont {P.}~\bibnamefont
  {Bhar}},\ }\bibfield  {title} {\bibinfo {title} {Dark energy stars in
  Tolman–Kuchowicz spacetime in the context of Einstein gravity},\ }\href
  {https://doi.org/https://doi.org/10.1016/j.dark.2021.100879} {\bibfield
  {journal} {\bibinfo  {journal} {Physics of the Dark Universe}\ }\textbf
  {\bibinfo {volume} {34}},\ \bibinfo {pages} {100879} (\bibinfo {year}
  {2021})}\BibitemShut {NoStop}%
\bibitem [{\citenamefont {Sharif}\ and\ \citenamefont
  {Waseem}(2020)}]{Sharif2020}%
  \BibitemOpen
  \bibfield  {author} {\bibinfo {author} {\bibfnamefont {M.}~\bibnamefont
  {Sharif}}\ and\ \bibinfo {author} {\bibfnamefont {A.}~\bibnamefont
  {Waseem}},\ }\bibfield  {title} {\bibinfo {title} {Impact of kuchowicz metric
  function on gravastars in $f(R,T)$ theory},\ }\href
  {https://doi.org/10.1140/epjp/s13360-020-00957-w} {\bibfield  {journal}
  {\bibinfo  {journal} {The European Physical Journal Plus}\ }\textbf {\bibinfo
  {volume} {135}} (\bibinfo {year} {2020})}\BibitemShut {NoStop}%
\bibitem [{\citenamefont {Ghosh}\ \emph {et~al.}(2021)\citenamefont {Ghosh},
  \citenamefont {Dey}, \citenamefont {Das}, \citenamefont {Chanda},\ and\
  \citenamefont {Paul}}]{Ghosh2021}%
  \BibitemOpen
  \bibfield  {author} {\bibinfo {author} {\bibfnamefont {S.}~\bibnamefont
  {Ghosh}}, \bibinfo {author} {\bibfnamefont {S.}~\bibnamefont {Dey}}, \bibinfo
  {author} {\bibfnamefont {A.}~\bibnamefont {Das}}, \bibinfo {author}
  {\bibfnamefont {A.}~\bibnamefont {Chanda}},\ and\ \bibinfo {author}
  {\bibfnamefont {B.~C.}\ \bibnamefont {Paul}},\ }\bibfield  {title} {\bibinfo
  {title} {Study of gravastars in rastall gravity},\ }\href
  {https://doi.org/10.1088/1475-7516/2021/07/004} {\bibfield  {journal}
  {\bibinfo  {journal} {Journal of Cosmology and Astroparticle Physics}\
  }\textbf {\bibinfo {volume} {2021}}\bibinfo  {number} { (07)},\ \bibinfo
  {pages} {004}}\BibitemShut {NoStop}%
\bibitem [{\citenamefont {Sharif}\ and\ \citenamefont
  {Saeed}(2021)}]{SHARIF2021}%
  \BibitemOpen
\bibfield  {number} {  }\bibfield  {author} {\bibinfo {author} {\bibfnamefont
  {M.}~\bibnamefont {Sharif}}\ and\ \bibinfo {author} {\bibfnamefont
  {M.}~\bibnamefont {Saeed}},\ }\bibfield  {title} {\bibinfo {title} {Study of
  gravastars admitting conformal motion in $f(\mathcal{R},\mathcal{T}^2)$
  gravity},\ }\href
  {https://doi.org/https://doi.org/10.1016/j.cjph.2021.08.009} {\bibfield
  {journal} {\bibinfo  {journal} {Chinese Journal of Physics}\ }\textbf
  {\bibinfo {volume} {701}},\ \bibinfo {pages} {388} (\bibinfo {year}
  {2021})}\BibitemShut {NoStop}%
\bibitem [{\citenamefont {Demorest}\ \emph {et~al.}(2010)\citenamefont
  {Demorest}, \citenamefont {Pennucci}, \citenamefont {Ransom}, \citenamefont
  {Roberts},\ and\ \citenamefont {Hessels}}]{Demorest2010ATN}%
  \BibitemOpen
  \bibfield  {author} {\bibinfo {author} {\bibfnamefont {P.}~\bibnamefont
  {Demorest}}, \bibinfo {author} {\bibfnamefont {T.}~\bibnamefont {Pennucci}},
  \bibinfo {author} {\bibfnamefont {S.}~\bibnamefont {Ransom}}, \bibinfo
  {author} {\bibfnamefont {M.}~\bibnamefont {Roberts}},\ and\ \bibinfo {author}
  {\bibfnamefont {J.}~\bibnamefont {Hessels}},\ }\bibfield  {title} {\bibinfo
  {title} {{Shapiro Delay Measurement of A Two Solar Mass Neutron Star}},\
  }\href {https://doi.org/10.1038/nature09466} {\bibfield  {journal} {\bibinfo
  {journal} {Nature}\ }\textbf {\bibinfo {volume} {467}},\ \bibinfo {pages}
  {1081} (\bibinfo {year} {2010})}\BibitemShut {NoStop}%
\bibitem [{\citenamefont {{Zeldovich}}(1972)}]{Zeldovich1972}%
  \BibitemOpen
  \bibfield  {author} {\bibinfo {author} {\bibfnamefont {Y.~B.}\ \bibnamefont
  {{Zeldovich}}},\ }\bibfield  {title} {\bibinfo {title} {{A hypothesis,
  unifying the structure and the entropy of the Universe}},\ }\href
  {https://doi.org/10.1093/mnras/160.1.1P} {\bibfield  {journal} {\bibinfo
  {journal} {\mnras}\ }\textbf {\bibinfo {volume} {160}},\ \bibinfo {pages}
  {1P} (\bibinfo {year} {1972})}\BibitemShut {NoStop}%
\bibitem [{\citenamefont {{Chandrasekhar}}(1964)}]{Chandrasekhar1964}%
  \BibitemOpen
  \bibfield  {author} {\bibinfo {author} {\bibfnamefont {S.}~\bibnamefont
  {{Chandrasekhar}}},\ }\bibfield  {title} {\bibinfo {title} {{The Dynamical
  Instability of Gaseous Masses Approaching the Schwarzschild Limit in General
  Relativity.}},\ }\href {https://doi.org/10.1086/147938} {\bibfield  {journal}
  {\bibinfo  {journal} {\apj}\ }\textbf {\bibinfo {volume} {140}},\ \bibinfo
  {pages} {417} (\bibinfo {year} {1964})}\BibitemShut {NoStop}%
\bibitem [{\citenamefont {Maurya}\ and\ \citenamefont
  {Maharaj}(2017)}]{Maurya2017}%
  \BibitemOpen
  \bibfield  {author} {\bibinfo {author} {\bibfnamefont {S.}~\bibnamefont
  {Maurya}}\ and\ \bibinfo {author} {\bibfnamefont {S.}~\bibnamefont
  {Maharaj}},\ }\bibfield  {title} {\bibinfo {title} {Anisotropic fluid spheres
  of embedding class one using karmarkar condition},\ }\href
  {https://doi.org/10.1140/epjc/s10052-017-4905-7} {\bibfield  {journal}
  {\bibinfo  {journal} {European Physical Journal C}\ }\textbf {\bibinfo
  {volume} {77}} (\bibinfo {year} {2017})}\BibitemShut {NoStop}%
\bibitem [{\citenamefont {Rej}\ \emph {et~al.}(2021)\citenamefont {Rej},
  \citenamefont {Bhar},\ and\ \citenamefont {Govender}}]{Rej2021}%
  \BibitemOpen
  \bibfield  {author} {\bibinfo {author} {\bibfnamefont {P.}~\bibnamefont
  {Rej}}, \bibinfo {author} {\bibfnamefont {P.}~\bibnamefont {Bhar}},\ and\
  \bibinfo {author} {\bibfnamefont {M.}~\bibnamefont {Govender}},\ }\bibfield
  {title} {\bibinfo {title} {Charged compact star in f
  $(\mathcal{R},\mathcal{T})$ gravity in tolman–kuchowicz spacetime},\ }\href
  {https://doi.org/10.1140/epjc/s10052-021-09127-3} {\bibfield  {journal}
  {\bibinfo  {journal} {European Physical Journal C}\ }\textbf {\bibinfo
  {volume} {81}},\ \bibinfo {pages} {316} (\bibinfo {year} {2021})}\BibitemShut
  {NoStop}%
\bibitem [{\citenamefont {Shamir}\ and\ \citenamefont
  {Zia}(2020)}]{Shamir2020b}%
  \BibitemOpen
  \bibfield  {author} {\bibinfo {author} {\bibfnamefont {M.~F.}\ \bibnamefont
  {Shamir}}\ and\ \bibinfo {author} {\bibfnamefont {S.}~\bibnamefont {Zia}},\
  }\bibfield  {title} {\bibinfo {title} {Gravastars in $f(R,G)$ gravity},\
  }\href {https://doi.org/10.1139/cjp-2019-0587} {\bibfield  {journal}
  {\bibinfo  {journal} {Canadian Journal of Physics}\ }\textbf {\bibinfo
  {volume} {98}},\ \bibinfo {pages} {849} (\bibinfo {year} {2020})}\BibitemShut
  {NoStop}%
\bibitem [{\citenamefont {Biswas}\ \emph {et~al.}(2020)\citenamefont {Biswas},
  \citenamefont {Shee}, \citenamefont {Guha},\ and\ \citenamefont
  {Ray}}]{Biswas2020}%
  \BibitemOpen
  \bibfield  {author} {\bibinfo {author} {\bibfnamefont {S.}~\bibnamefont
  {Biswas}}, \bibinfo {author} {\bibfnamefont {D.}~\bibnamefont {Shee}},
  \bibinfo {author} {\bibfnamefont {B.~K.}\ \bibnamefont {Guha}},\ and\
  \bibinfo {author} {\bibfnamefont {S.}~\bibnamefont {Ray}},\ }\bibfield
  {title} {\bibinfo {title} {{Anisotropic strange star with
  Tolman\textendash{}Kuchowicz metric under $f(R,T)$ gravity}},\ }\href
  {https://doi.org/10.1140/epjc/s10052-020-7725-0} {\bibfield  {journal}
  {\bibinfo  {journal} {Eur. Phys. J. C}\ }\textbf {\bibinfo {volume} {80}},\
  \bibinfo {pages} {175} (\bibinfo {year} {2020})}\BibitemShut {NoStop}%
\bibitem [{\citenamefont {Majid}\ and\ \citenamefont
  {Sharif}(2020)}]{Majid2020}%
  \BibitemOpen
  \bibfield  {author} {\bibinfo {author} {\bibfnamefont {A.}~\bibnamefont
  {Majid}}\ and\ \bibinfo {author} {\bibfnamefont {M.}~\bibnamefont {Sharif}},\
  }\bibfield  {title} {\bibinfo {title} {{Quark Stars in Massive
  Brans\textendash{}Dicke Gravity with Tolman\textendash{}Kuchowicz
  Spacetime}},\ }\href {https://doi.org/10.3390/universe6080124} {\bibfield
  {journal} {\bibinfo  {journal} {Universe}\ }\textbf {\bibinfo {volume} {6}},\
  \bibinfo {pages} {124} (\bibinfo {year} {2020})}\BibitemShut {NoStop}%
\bibitem [{\citenamefont {Jasim}\ \emph {et~al.}(2018)\citenamefont {Jasim},
  \citenamefont {Deb}, \citenamefont {Ray}, \citenamefont {Gupta},\ and\
  \citenamefont {Chowdhury}}]{Jasim2018}%
  \BibitemOpen
  \bibfield  {author} {\bibinfo {author} {\bibfnamefont {M.~K.}\ \bibnamefont
  {Jasim}}, \bibinfo {author} {\bibfnamefont {D.}~\bibnamefont {Deb}}, \bibinfo
  {author} {\bibfnamefont {S.}~\bibnamefont {Ray}}, \bibinfo {author}
  {\bibfnamefont {Y.~K.}\ \bibnamefont {Gupta}},\ and\ \bibinfo {author}
  {\bibfnamefont {S.~R.}\ \bibnamefont {Chowdhury}},\ }\bibfield  {title}
  {\bibinfo {title} {{Anisotropic strange stars in Tolman\textendash{}Kuchowicz
  spacetime}},\ }\href {https://doi.org/10.1140/epjc/s10052-018-6072-x}
  {\bibfield  {journal} {\bibinfo  {journal} {Eur. Phys. J. C}\ }\textbf
  {\bibinfo {volume} {78}},\ \bibinfo {pages} {603} (\bibinfo {year}
  {2018})}\BibitemShut {NoStop}%
\bibitem [{\citenamefont {Farasat~Shamir}\ and\ \citenamefont
  {Naz}(2020)}]{FarasatShamir2020}%
  \BibitemOpen
  \bibfield  {author} {\bibinfo {author} {\bibfnamefont {M.}~\bibnamefont
  {Farasat~Shamir}}\ and\ \bibinfo {author} {\bibfnamefont {T.}~\bibnamefont
  {Naz}},\ }\bibfield  {title} {\bibinfo {title} {{Stellar structures in
  $f(\mathcal{G})$ gravity with Tolman\textendash{}Kuchowicz spacetime}},\
  }\href {https://doi.org/10.1016/j.dark.2020.100472} {\bibfield  {journal}
  {\bibinfo  {journal} {Phys. Dark Univ.}\ }\textbf {\bibinfo {volume} {27}},\
  \bibinfo {pages} {100472} (\bibinfo {year} {2020})}\BibitemShut {NoStop}%
\end{thebibliography}
\end{document}